\documentclass[twocolumn,aps,floats]{revtex4}
\usepackage{epsfig,graphicx}
\usepackage{amssymb}

\newcommand{\beq}{\begin{equation}}
\newcommand{\eeq}{\end{equation}}
\newcommand{\be}{\begin{equation}}
\newcommand{\ee}{\end{equation}}
\newcommand{\ba}{\begin{array}}
\newcommand{\ea}{\end{array}}
\newcommand{\beqa}{\begin{eqnarray}}
\newcommand{\eeqa}{\end{eqnarray}}
\newcommand{\bea}{\begin{eqnarray}}
\newcommand{\eea}{\end{eqnarray}}

\newcommand{\gsim}{\stackrel{>}{_\sim}}

\newcommand{\cA}{{\cal A}}
\newcommand{\cO}{{\cal O}}
\newcommand{\cB}{{\cal B}}
\newcommand{\BR}{{\cal B}}
\newcommand{\cL}{{\cal L}}

\newcommand{\no}{\nonumber}

\newcommand{\Btaun}{{B \to \tau \nu}}

\newcommand{\kpn}{K^+\to\pi^+\nu\bar\nu}
\newcommand{\klpn}{K_L\to\pi^0\nu\bar\nu}

\def\npb#1#2#3{    {\it Nucl. Phys.}~B {\bf #1}, #3 (#2)}
\def\plb#1#2#3{    {\it Phys. Lett.}~B {\bf #1}, #3 (#2)}

\def\prd#1#2#3{    {\it Phys. Rev.}~D {\bf #1}, #3 (#2)}

\def\prep#1#2#3{   {\it Phys. Rep. }{\bf #1}, #3 (#2)}
\def\prl#1#2#3{    {\it Phys. Rev. Lett. }{\bf #1}, #3 (#2)}
\def\ptp#1#2#3{    {\it Prog. Theor. Phys. }{\bf #1}, #3 (#2)}

\def\epjc#1#2#3{   {\it Eur. Phys. J.}~C {\bf #1}, #3 (#2)}

\begin{document}

\title{Flavour Physics: Now and in the LHC era\footnote{~To appear in the
the proceedings of {\em Lepton Photon 2007} (Daegu, Korea, Aug.~13-18 2007). } }

\author{Gino Isidori}

\affiliation{Scuola Normale Superiore and INFN, Piazza dei Cavalieri 7, 
I-56126 Pisa, Italy \\
INFN, Laboratori Nazionali di Frascati, Via E.Fermi 40, I-00044 Frascati, Italy}

\begin{abstract}
We present an overview of 
what we learned so far from low-energy flavour 
observables, concerning physics beyond the 
Standard Model, and what we could still learn from 
further studies in flavour physics in the next few years.
\end{abstract}

\maketitle


\section{Introduction:  the main lessons of Flavour Physics}

\begin{figure}[t]
\begin{center}
\includegraphics[width=78mm]{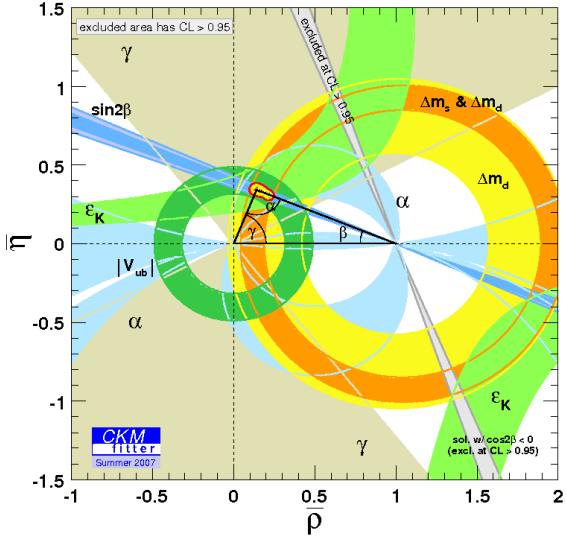}
\caption{\label{fig:fits} Fit of the CKM unitarity triangle 
within the SM~\cite{Charles:2004jd} (see also~\cite{Bona:2007vi}).}
\end{center}
\end{figure}

In the last few years there has been a great 
experimental progress in quark and lepton 
flavour physics. In the quark sector, 
the validity of the Standard Model (SM)
has been strongly reinforced by a 
series of challenging tests. As summarised by the
plots shown in Fig.~\ref{fig:fits}, all the relevant 
SM parameters controlling quark-flavour dynamics (the quark masses and 
the angles of the Cabibbo-Kobayashi-Maskawa matrix \cite{CKM}) 
have been determined with good accuracy.
More important, several suppressed observables
(such as $\Delta M_{B_d}$, $\Delta M_{B_s}$, $\cA^{_{\rm CP}}_{K \Psi}$,
$B\to X_s \gamma$, $\epsilon_K$, \ldots) potentially sensitive to 
New Physics (NP) have been measured with good accuracy,
showing no deviations from the SM.
The situation is somehow similar to the flavour-conserving 
electroweak precision observables (EWPO) after LEP: 
the SM works very well and genuine 
one-loop electroweak effects have been tested with 
relative accuracy in the $10\%$--$30\%$ range.
Similarly to the EWPO case, also in the quark flavour 
sector NP effects can only appear as a small correction 
to the leading SM contribution.

The situation of the lepton sector is more uncertain 
but also more exciting. The discovery of neutrino 
oscillations has two very significant implications:
i) the SM is not complete; ii) there exists new 
flavour structures in addition to the three 
SM Yukawa couplings. We have not yet enough information
to unambiguously determine how the SM Lagrangian 
should be modified in order to describe the 
phenomenon of neutrino oscillations. However, 
natural explanations point toward the 
existence of new degrees of freedom 
with explicit breaking of lepton number
at very high energy scales, 
in agreement with the expectations of 
Grand Unified Theories (GUT).

If the SM is not a complete theory, it is natural to expect new degrees
of freedom around or slightly above the electroweak scale 
(the energy domain that will 
be fully explored for the first time at the LHC). Indeed 
we cannot extend the validity of the SM above the TeV range 
without a serious fine-tuning problem in the Higgs sector
(see e.g.~Ref.~\cite{BS}).
In constructing a realistic SM extension we should then try to 
reconcile three apparently conflicting requirements:
\begin{enumerate}
\item[i.] new degrees of freedom around the electroweak scale,
\item[ii.] no significant deviations for the SM in the quark sector 
           (as well as no significant effects in EWPO);
\item[iii.] non-standard flavour structures in the lepton sector.
\end{enumerate}
The rest of this talk is devoted to discuss how these three
points can be reconciled, and why they imply 
that a few specific measurements in the 
flavour sector will still be very interesting 
also in the LHC era.

\section{What we learned so far about New Physics}

We can follow three main strategies to describe and quantify 
what we learned so far about NP from quark-flavour observables.

\medskip
{\bf I.} {\em Generic EFT approach.} \\
As long as we are interested in processes occurring well below 
the electroweak scale (such as $B$, $D$ and $K$ 
decays), we can integrate out the new degrees of freedom 
and describe NP effects --in full generality-- by means 
of an Effective Field Theory (EFT) approach.
The SM Lagrangian becomes the renormalizable part of a more general 
local Lagrangian which includes an infinite tower of higher-dimensional 
operators, constructed in terms of SM fields and 
suppressed by inverse powers of a scale $\Lambda_{\rm NP} > v = 174$~GeV. 
This general bottom-up approach allows us to analyse all realistic 
extensions of the SM in terms of a limited number of parameters 
(the coefficients of the higher-dimensional operators). 
The disadvantage of this strategy is that  
it does not allow us to establish correlations 
of New Physics (NP) effects at low and high energies (the scale
$\Lambda_{\rm NP}$ defines the cut-off of the EFT). 
The number of correlations among different low-energy 
observables is also very limited, unless some restrictive 
assumptions about the structure of the EFT are employed. 

\medskip
{\bf II.} {\em Explicit NP models.} \\
The generic EFT approach is somehow the 
opposite of the standard top-bottom strategy toward NP, 
where a given theory --and a specific set of parameters-- 
are employed to evaluate possible deviations from the SM.  
The top-bottom approach usually allows to establish several correlations,
both a low-energies and between low- and high-energy observables. 
However, the price to pay is the loss of generality. This is 
quite a high price given our limited knowledge about the physics 
above the electroweak scale. 

\medskip
{\bf III.} {\em EFT with explicit flavour symmetries.} \\
An interesting compromise between these two extreme strategies 
is obtained implementing specific symmetry restrictions 
on the EFT. The extra constraints increase the number of 
correlations in low-energy observables. The experimental 
tests of such correlations allows us to test/establish 
general features of the NP model (possibly valid both 
at low- and high-energies). In particular, 
$B$, $D$ and $K$ decays are extremely useful in determining 
the flavour-symmetry breaking pattern of the NP model. 
The EFT approaches based on the Minimal Flavour Violation 
(MFV) hypothesis and its variations 
(MFV at large $\tan\beta$, n-MFV,~\ldots)
have exactly this goal.

\subsection{Generic EFT approaches and the flavour problem}
\label{sect:flav_prob}

\begin{figure*}[t]
\begin{center}
\includegraphics[width=75mm]{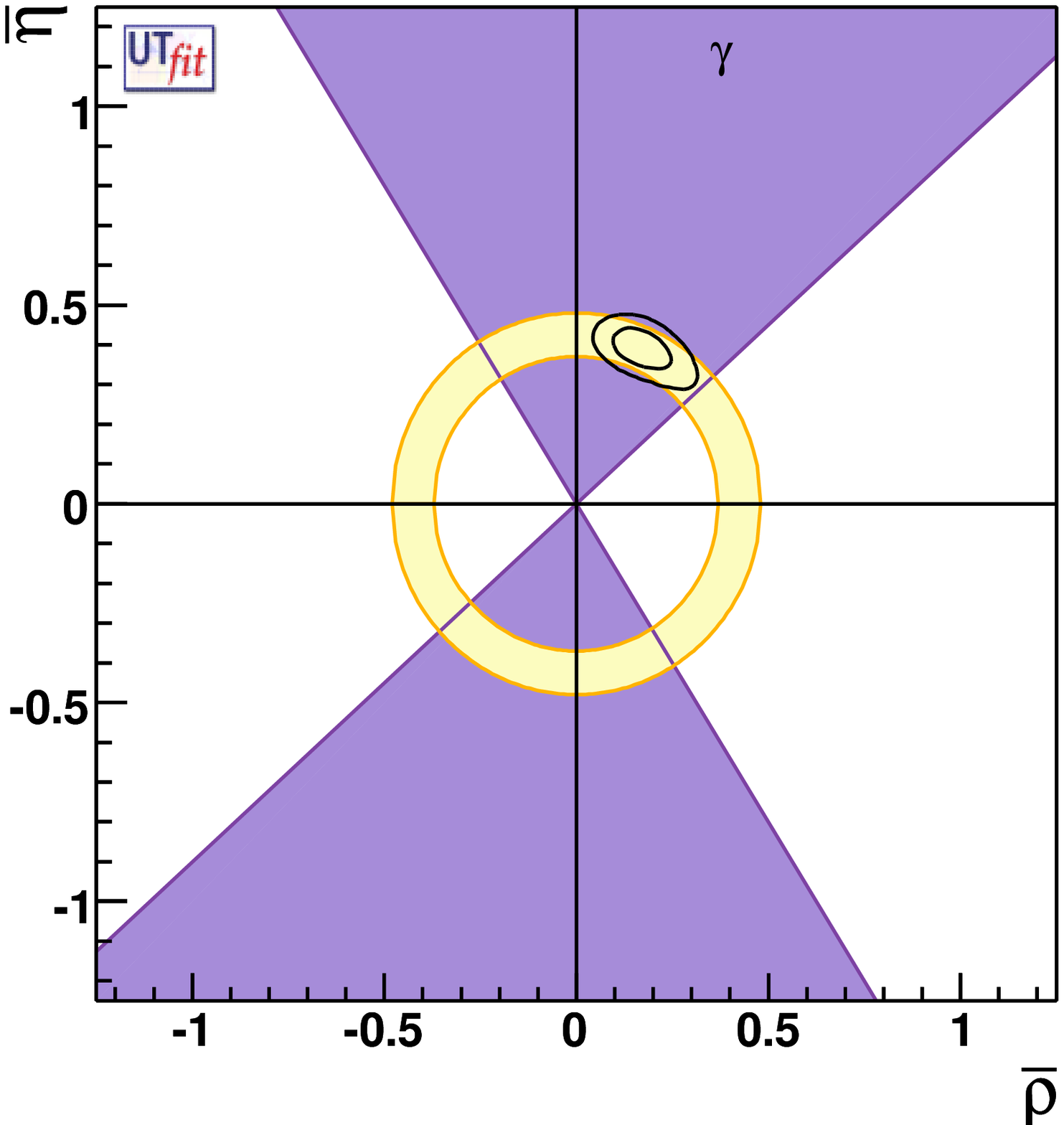}
\includegraphics[width=75mm]{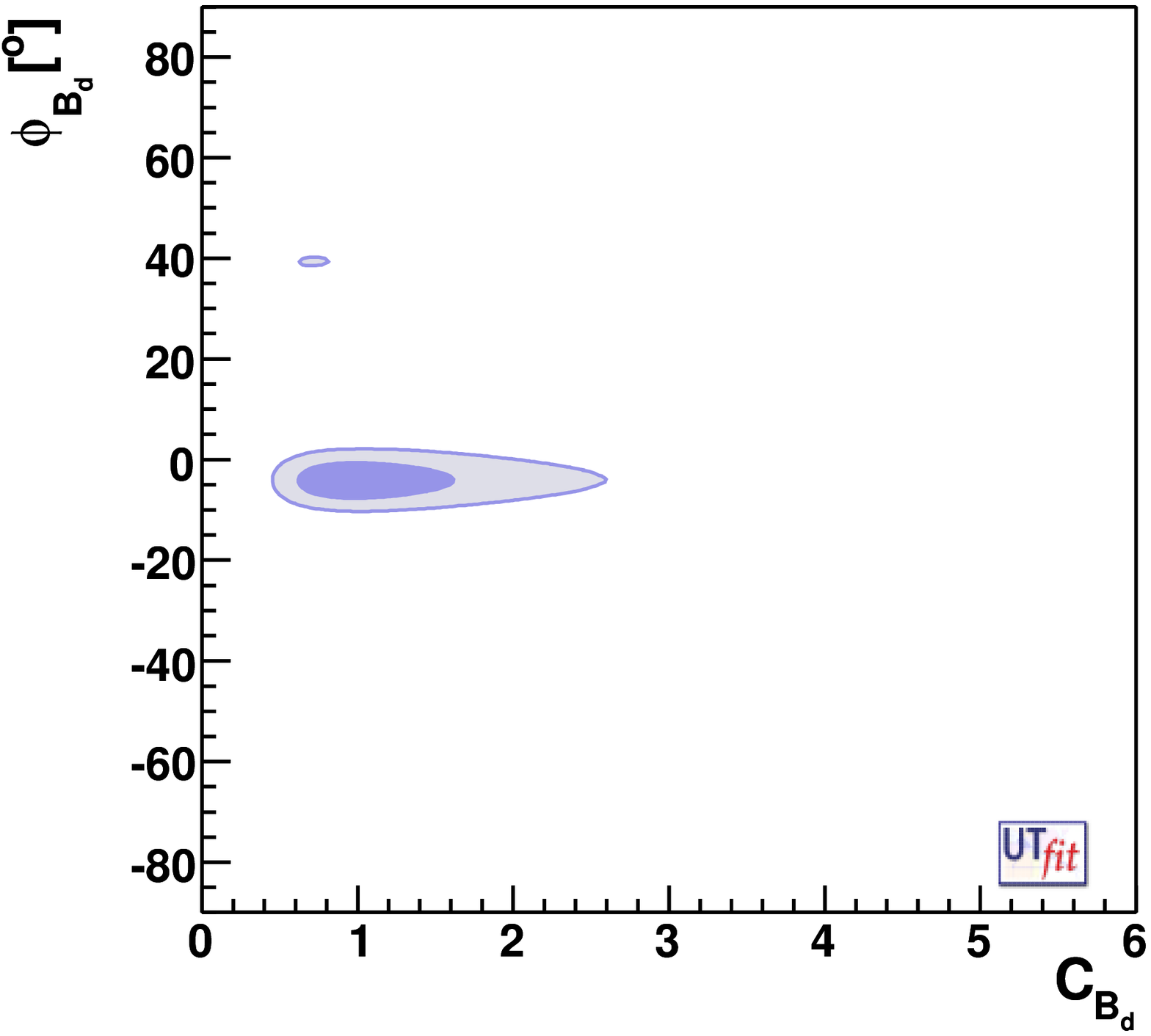}
\caption{Left: Constraints on the $\bar \rho$--$\bar \eta$ plane 
using tree-level observables only. Right:
Constraints on the effective parameters encoding NP effects 
in the $B_d$--$\bar B_d$ mixing amplitude (magnitude and phase) 
\cite{Bona:2007vi}.}
\label{fig:npCphi}
\end{center}
\end{figure*}

The NP contributions to the higher-dimensional operators of the 
EFT  should naturally induce large effects 
in processes which are not mediated by tree-level SM amplitudes, 
such as meson-antimeson mixing ($\Delta F=2$ amplitudes)
or flavour-changing neutral-current (FCNC) rare decays.
On the other hand, it is usually a good approximation to neglect 
non-standard effects in processes which are mediated by 
tree-level SM amplitudes. A general analyses of $\Delta F=2$
observables based on the latter assumption has recently 
been performed by the UTfit Collaboration~\cite{Bona:2007vi}
(earlier studies can be found also in Ref.~\cite{Charles:2004jd}).
The results are summarised by the plots in Fig.~\ref{fig:npCphi}. 

First of all, it is interesting to note that present data, 
in particular the determination of $\gamma$ and $|V_{ub}|$, 
allow a rather precise determination of the CKM matrix
using tree-level processes only (Fig.~\ref{fig:npCphi} left).
This allows a model-independent comparison of the experimental 
data on meson-antimeson mixing   
with the corresponding theoretical SM predictions.
NP effects in $\Delta F=2$ amplitudes can simply be 
parametrized in terms of a modulo and a phase for 
each  meson-antimeson amplitude, 
\begin{equation} 
\frac{\langle
    M |H_\mathrm{eff}^\mathrm{full}|\bar{M} \rangle} {\langle
    M |H_\mathrm{eff}^\mathrm{SM}|\bar{M} \rangle}=
    C_{M}  e^{2 i \phi_{M}}
    \label{eq:paranp}
\end{equation}
such that the SM is recovered for $C_{M}=1$ and $\phi_{M}=0$.
The main conclusions which can be drawn form the 
present analyses can be summarized as follows: 
\begin{itemize}
\item 
In all the three accessible amplitudes       
($K^0$--$\bar K^0$, $B_d$--$\bar B_d$, and $B_s$--$\bar B_s$)
the magnitude of the new-physics amplitude cannot exceed, in size, 
the SM short-distance contribution. The latter is suppressed both 
by the GIM mechanism and by the hierarchical structure of the 
CKM matrix ($|V_{td}|, |V_{ts}| \ll 1$):
\bea
&& \cA^{\Delta F=2}_{\rm SM} \sim  
   \frac{ G_F^2 M_W^2 }{2 \pi^2} \left(V_{ti}^* V_{tj} \right)^2\times \no \\
&& \quad  \times\langle \bar  M |  (\bar Q_L^i \gamma^\mu Q_L^j )^2  | M \rangle
\qquad (i,j = d,s)\ 
\eea
Therefore, new-physics models with TeV-scale flavored 
degrees of freedom and $\cO(1)$ flavour-mixing couplings are 
essentially ruled out: denoting by $C_{ij}$ the 
flavour-mixing coupling in the NP model,
\be
  \cA^{\Delta F=2}_{\rm NP} \sim  \frac{C_{ij}}{2 \Lambda^2} 
  \langle \bar  M | (\bar Q_L^i \gamma^\mu Q_L^j )^2  | M \rangle  
\label{eq:tre} 
\ee
the condition $| \cA^{\Delta F=2}_{\rm NP}| <  | \cA^{\Delta F=2}_{\rm SM} |$
implies
\bea
\Lambda < \frac{ 3.4~{\rm TeV} }{| V_{ti}^* V_{tj}|/|C_{ij}|^{1/2}  }
<  \left\{ \ba{l}  
9\times 10^3~{\rm TeV} \times |C_{sd}|^{1/2} \!\!\!\!\!\!\! \\ 
4\times 10^2~{\rm TeV} \times |C_{bd}|^{1/2} \!\!\!\!\!\!\! \\
7\times 10^1~{\rm TeV} \times |C_{bs}|^{1/2} \!\!\!\!\!\!\! \ea \!\!\!\!\!\!\!\!\!\!
\right. \no
\eea
\item 
As clearly shown in Fig.~\ref{fig:npCphi}, in the 
$B_d$--$\bar B_d$ case there is still room 
for a new-physics contribution up to $\sim 50\%$
of the SM one ($C_{B_d}$ can be substantially 
different from unity). However, this is possible only 
if the new-physics contribution is aligned in phase 
with respect to the SM amplitude ($\phi_{B_d}$ close to zero). 
A similar conclusion holds also for the $K^0$--$\bar K^0$
amplitude.
\item
Contrary to $B_d$--$\bar B_d$ and $K^0$--$\bar K^0$
amplitudes, at present there is only a very  loose bound on the 
CPV phase of the  $B_s$--$\bar B_s$ mixing amplitude. 
This leaves open the possibility of observing a large 
$\cA_{\rm CP}(B_s \to J/\Psi \phi)$ at LHCb, which 
would be a clear signal of physics beyond the SM.
\end{itemize}
The strong bounds on $\Lambda$ in models with
generic flavour structure ($C_{ij}\sim 1$) is a manifestation 
of what in many specific frameworks (supersymmetry, technicolour, etc.)
goes under the name of {\em flavour problem}:
if we insist with the theoretical prejudice that new physics has to 
emerge in the TeV region, we have to conclude that the new theory 
possesses a highly non-generic flavour structure. 
Interestingly enough, this structure has not been clearly identified yet,
mainly because the SM, i.e.~the low-energy 
limit of the new theory, doesn't possess an exact flavour symmetry.

The most reasonable (but also most {\em pessimistic}) 
solution to the flavour problem is the so-called 
Minimal Flavour Violation hypothesis~\cite{Georgi,MFV,MLFV,MFVBB}. 
Under this assumption, which will be discussed below, 
the first two items listed above find a natural
explanation.

\subsection{Minimal Flavour Violation}

The main idea of MFV is that flavour-violating 
interactions are linked to the
known structure of Yukawa couplings also beyond the SM. 
As a result, non-standard contributions in FCNC 
transitions turn out to suppressed to a level consistent 
with experiments even for $\Lambda \sim$~few TeV.
On the most interesting aspects of the MFV hypothesis 
is that it can easily be implemented within the 
general EFT approach to new physics~\cite{MFV,MLFV}. 
This allows us to establish general and unambiguous correlations 
among NP effects in various rare decays. 
These falsifiable predictions are a key ingredient   
to identify in a model-independent way the flavour structure 
of the new-physics model.

In a more quantitative way, the MFV construction consists 
in identifying the flavour symmetry and symmetry-breaking structure 
of the SM and enforce it the EFT. 
In the quark sector this procedure is unambiguous:
the largest group of flavour-changing field 
transformations commuting with the gauge group is 
${\cal G}_q = SU(3)_{Q_L}\times SU(3)_{U_R}\times SU(3)_{D_R}$, 
and this group is broken only by the two 
$3\times \bar 3$ structures of the Yukawa interaction:
\bea
\cL_{\rm Y}^{\rm quark}  &=&  {\bar Q}^i_L (Y_U)_{ij} U^j_R  H_U  + \no \\ 
&& + {\bar Q}_L^i (Y_D)_{ij} D^j_R  H_D +{\rm h.c.} \quad 
\label{eq:LYtree0}
\eea 
The invariance of the SM Lagrangian under 
${\cal G}_q$ can be formally recovered elevating the Yukawa matrices to
spurion fields with appropriate transformation properties under
${\cal G}_q$. The hypothesis of MFV states that these are the only 
spurions breaking ${\cal G}_q$ also beyond the SM. Within the  
effective theory formulation, this implies that all the higher dimensional 
operators constructed from SM and Yukawa fields  must be 
(formally) invariant under ${\cal G}_q$. 

It is then easy to realize that, similarly to the pure SM case, 
the leading coupling ruling all FCNC transitions 
with external down-type quarks is  
$(Y_U Y_U^\dagger)_{ij}\approx y_t^2  V^*_{3i} V_{3j}$,
with $y_t =m_t/v \approx 1$ (in the down-quark mass-eigenstate basis). 
As a result, within this framework the 
coefficients of the higher-dimensional operators have the  
same CKM suppression of the corresponding SM amplitudes
and the bounds on the new-physics scale are 
in the few TeV range. This is already clear 
from~Eq.(\ref{eq:tre}), once we set $C_{ij} = y_t^2 V^*_{3i} V_{3j}$;
statistically well defined and updated bounds can be in Ref.~\cite{Bona:2007vi}. 
Moreover, the flavour structure of $Y_U Y_U^\dagger$ implies a well-defined link among 
possible deviations from the SM in FCNC transitions 
of the type $s\to d$, $b\to d$, and  
$b\to s$ (the only quark-level transitions where 
observable deviations from the SM are expected).

\begin{figure*}[t]
\begin{center}
\includegraphics[width=75mm]{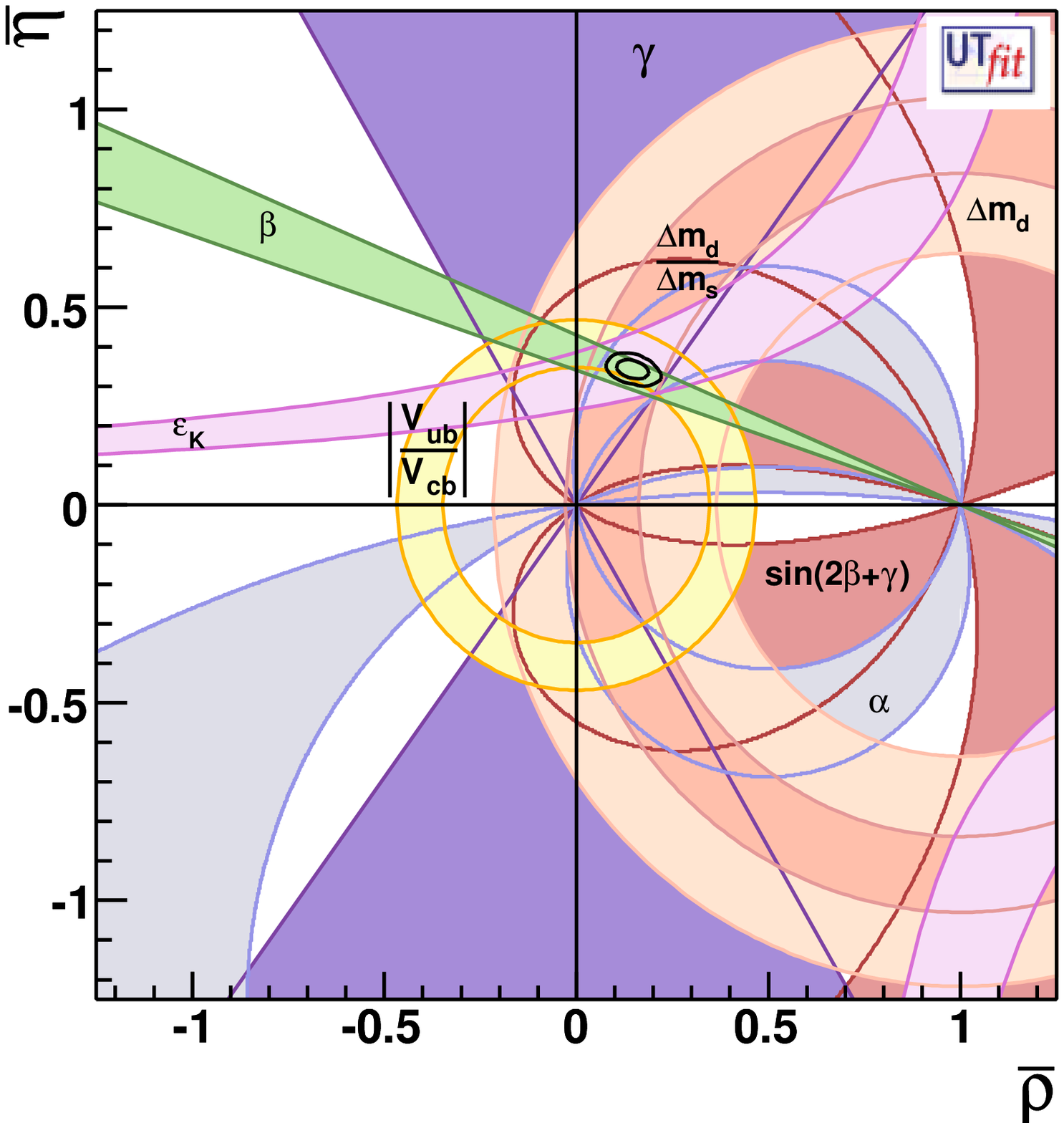}
\includegraphics[width=75mm]{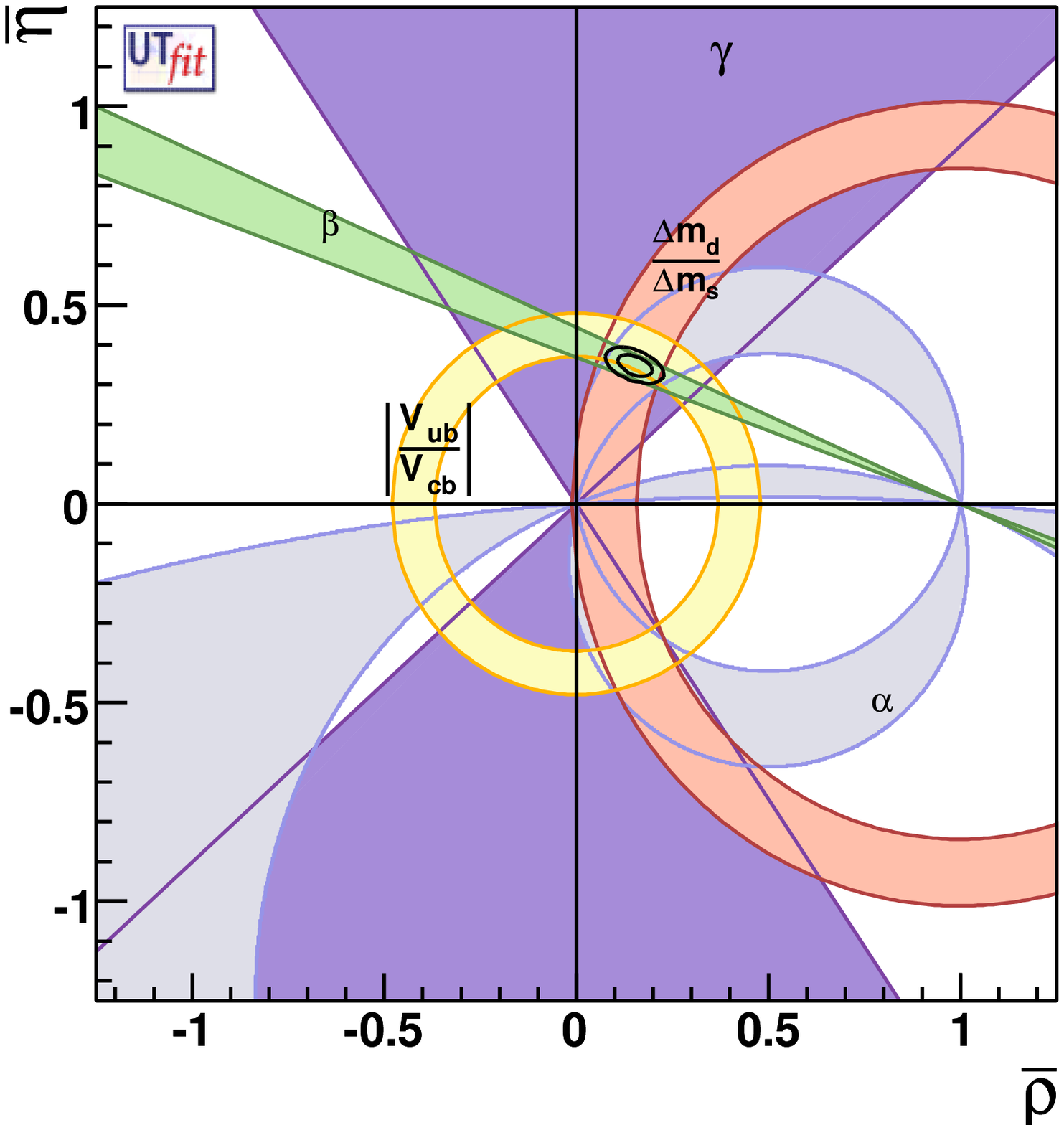}
\caption{\label{fig:UTfits} Fit of the CKM unitarity 
triangle within the SM (left) and 
in generic extensions of the SM satisfying the MFV hypothesis 
(right)~\cite{Bona:2007vi}. }
\end{center}
\end{figure*}
  
The idea that the CKM matrix rules the strength of 
FCNC transitions also beyond the SM is a concept
that has been implemented and discussed in several works, 
especially after the first results of the $B$ factories
(see e.g.~Ref.~\cite{MFV2}). 
However, it is worth stressing that the CKM matrix 
represents only one part of the problem: a key role in
determining the structure of FCNCs is also played  by quark masses
(via the GIM mechanism), or by the Yukawa eigenvalues. 
In this respect, the above MFV criterion provides the maximal protection 
of FCNCs (or the minimal violation of flavour symmetry), since the full 
structure of Yukawa matrices is preserved. Moreover, 
contrary to other approaches, the above MFV criterion 
is based on a renormalization-group-invariant symmetry argument,
which can easily be extended to TeV-scale 
effective theories where new degrees
of freedoms, such as extra scalar fields (see~e.g.~\cite{MW})
or SUSY partners of the SM fields~(see 
e.g.~\cite{Altmannshofer:2007cs,Nikolidakis:2007fc}), 
are included. Finally, this symmetry and
symmetry-breaking pattern can explicitly be implemented in 
well-motivated UV completions of the SM valid up to very high 
energy scales (see e.g.~\cite{MFV_expl1,MFV_expl2}).

As shown in Fig.~\ref{fig:UTfits}, the MFV hypothesis provides a
natural (a posteriori) justification of why no NP effects have 
been observed in the quark sector: by construction, most of the clean 
observables measured at $B$ factories are insensitive to NP effects 
in this framework. However, it should be stressed that we are still
very far from having proved the validity of this hypothesis 
from data. A proof of the MFV hypothesis can be achieved 
only with a positive evidence of physics beyond 
the SM exhibiting the flavour pattern (link between $s\to d$, $b\to d$, and  
$b\to s$) predicted by the MFV assumption~\cite{MFV}.
So far we have only bounds on NP effects in the flavour sector,
and it could well be that the new theory includes 
non-minimal sources of flavour symmetry breaking with specific 
flavour structures, such as those discussed in Ref.~\cite{NMFV}. 
It is also conceivable that there is not an underlying 
flavour symmetry, and the suppression of FCNCs is
of dynamical origin. This happens for instance in scenarios 
with hierarchical fermion wave functions~\cite{GSS}, which are well 
motivated by models with warped extra-dimensions~\cite{Warped}.

Last but not least, it is worth to stress that 
even within the pessimistic MFV framework the lepton sector 
could still be very exciting. The implementation of the 
MFV hypothesis in the lepton sector is not as 
straightforward as for the quark sector~\cite{MLFV}. 
But if the breaking of lepton flavour and total lepton number 
are decoupled, rare LFV decays such as $\mu\to e\gamma$ 
could be within the reach of the next generation of 
experiments even in a MFV framework~\cite{MLFV}.

\subsection{Flavour constraints in explicit models}

In all explicit NP scenarios the constraints
of flavour physics play a very important role.
This is obvious in cases where the model allow the 
existence of new sources of flavour symmetry breaking.
A typical example is the MSSM with generic flavour 
structures~\cite{Gabbiani:1996hi}: 
here each flavour observable is used to set a limit 
on a specific combination of non-diagonal entries of the 
sfermion mass matrices (see e.g.~Ref.~\cite{Jager:2007bm} 
for a recent discussion).

The importance of  flavour observables is less obvious 
in constrained models, such as MSSM scenarios with MFV. 
The situation here turns out to be even more interesting
than in generic models: the number of  free parameters
is substantially reduced and a given observable put
constrains which are relevant for several other processes
(even beyond the  flavour sector).
As a result, the consistency of the model is probed to a deep level. 

An illustration of this fact in the context of the mSUGRA scenario 
has been presented in Ref.~\cite{Heinemeyer:2007cn}:
the information derived by $B\to X_s \gamma$ poses a significant 
constraint on the model, which is compatible with those 
derived from flavour conserving processes. In particular, the heavy 
stop mass required by  $B\to X_s \gamma$ is one of the main 
ingredients which pushes the mass of the 
light Higgs boson above the LEP bound~\cite{Buchmueller:2007zk}.

As recently shown in Ref.~\cite{Albrecht:2007ii}, there are also 
specific supersymmetric MFV frameworks which are essentially 
ruled out by the recent results of 
flavour physics. In particular, the present 
constraints from $\BR(B\to\tau\nu)$, $\BR(B_s\to\mu^+\mu^-)$
and $\BR(B\to X_s \gamma)$,
puts in serious difficulties the $SO(10)$ GUT model of 
Dermisek and Raby~\cite{Dermisek:2005ij}, 
a specific example of MFV scenario with large $\tan\beta$.

\section{Flavour Physics in the LHC era}

If new particles or, more generally, new degrees of freedom,  are present 
in the TeV energy range, there are good chances that part of them will 
be discovered at the LHC. This does not mean that the complete
structure of the new model can easily be determined at the LHC: the direct 
discovery of new particles is only one of the ingredients necessary to
achieve this goal. As already discussed in the previous section, 
some of the parameters of the model (in particular its flavour structure)
can only be determined with improved measurements in 
the flavour sector. A brief survey of the most interesting low-energy 
flavour observables in this perspective, focusing on MSSM 
scenarios with MFV (or approximate MFV), is presented in the following.

\subsection{Helicity-suppressed observables and the large $\tan\beta$ scenario}

The Higgs sector of the MSSM consists of two  $SU(2)_L$ scalar doublets, 
coupled separately to up- and down-type quarks
\bea
\cL^{\rm tree}_{\rm H}  &=&  {\bar Q}_L {Y_U} U_R  H_U  
+ {\bar Q}_L Y_D D_R  H_D + \no \\ 
&& + {\bar L}_L {Y_E} E_R  H_D  + V(H_U,H_D)+{\rm h.c.} \quad 
\label{eq:LYtree}
\eea
A key parameter of this sector is the ratio 
of the two Higgs vevs:
$\tan\beta = \langle H_U \rangle /\langle H_D \rangle$.
Varying  $\tan\beta$ leads to modify the overall normalization 
of the two Yukawa couplings and, for $\tan\beta \sim 40$--$50$, 
we can achieve the interesting unification of top 
and bottom Yukawa couplings.

The variation of
$\tan\beta$ do not change the misalignment in fla\-vour space
of the two Yukawa couplings. This implies that flavour-changing 
observables not suppressed by powers of down-type 
quark masses (i.e.~most of the experimentally accessible 
observables) are not sensitive to the value of $\tan\beta$. 
If the model has a MFV structure, the
phenomenological consequences of  $\tan\beta \gg 1$
show up only in the few observables sensitive to 
helicity-suppressed amplitudes.
These are confined to the $B$-meson system
(because of the large $b$-quark Yukawa coupling),
with the notable exception of $K\to\ell\nu$ decays.
We can divide the most interesting observables
in three classes: the charged-current processes $B(K) \to \ell \nu$,
the rare decays $B_{s,d} \to \ell^+\ell^-$, 
and the FCNC transition $B \to X_s \gamma$. 

It is worth to stress that, beside the theoretical interest,
the large $\tan\beta$  regime of the MSSM could also 
provide a natural explanation of the $a_\mu = (g-2)_\mu/2$ anomaly, 
which is now a solid $3\sigma$ effect:
$ \Delta a_{\mu} =  a_{\mu}^{\rm exp} - a_{\mu}^{\rm SM}  
\approx (2.9 \pm 0.9) \times 10^{-9}$~\cite{gm2}.
The size of this discrepancy is large 
compared to the electroweak SM contribution
($\Delta a_{\mu}^{\rm e.w.} \approx 1.5 \times 10^{-9}$). This 
large discrepancy can easily be explained 
by the fact that $a_\mu$ is a(flavour-conserving) 
helicity suppressed observable, whose non-standard 
contribution can be enhanced compared to the 
SM one by increasing the value of $\tan\beta$:
\be 
\Delta a^{\rm MSSM}_\mu \approx  \tan\beta  \times 
\Delta a_{\mu}^{\rm e.w.} \times 
\left( \frac{M_W}{\widetilde M_{\rm slept}} \right)^2
\ee
For values of $\tan\beta \gsim 10$ the $M_W/{\widetilde M_{\rm slept}}$
suppression can easily be compensated for sleptons 
well above the $W$ mass, in prefect agreement with 
the constraints of electroweak precision tests.

\subsubsection{$B(K) \to \ell \nu$}

The charged-current processes $P \to \ell \nu$ 
are the simplest flavour-violating 
helicity suppressed observables. Here both SM and 
Higgs-mediated contributions (sensitive to $\tan\beta$) 
appear already at the tree level. The $H^{\pm}$ contribution 
is proportional to the Yu\-ka\-wa couplings of quarks and leptons, but
it can compete with the $W^{\pm}$ exchange
thanks to the helicity suppression of 
$P \to \ell \nu$~\cite{Hou}.  
Taking into account the resummation of the leading $\tan\beta$
corrections to all orders, the $H^\pm$ contributions to 
the  $P \to \ell \nu $ amplitude within a MFV supersymmetric framework 
leads to the following ratio~\cite{IP,btnu}:
\bea
 R_{P\ell\nu} &=& \frac{\BR( P\ell\nu)}{\BR^{\rm SM}( P\ell\nu)} \no \\
& \quad \stackrel{\rm SUSY}{=}& \left[1-\left(\frac{m^{2}_P}{m^{2}_{H^\pm}}\right)
\frac{\tan^2\beta}{(1+\epsilon_0\tan\beta)}
\right]^2\quad
\label{eq:Btn}
\eea
where $\epsilon_0$ denotes the effective coupling which 
parame\-trizes the non-holomorphic corrections to 
the down-type Yukawa interaction~\cite{Babu,IR}. 
For a natural choice of the MSSM parameters, 
Eq.~(\ref{eq:Btn}) implies a suppression 
with respect to the SM in $B$ decays of
few$\times10\%$ (but an enhancement is also possible 
for very light $M_{H^\pm}$) and an effect 100
times smaller in $K$ decays (where the branching ratio
is necessarily smaller than $\BR^{\rm SM}$). 

\begin{figure}[t]
\centering
\hskip -1 cm
\includegraphics[scale=0.35]{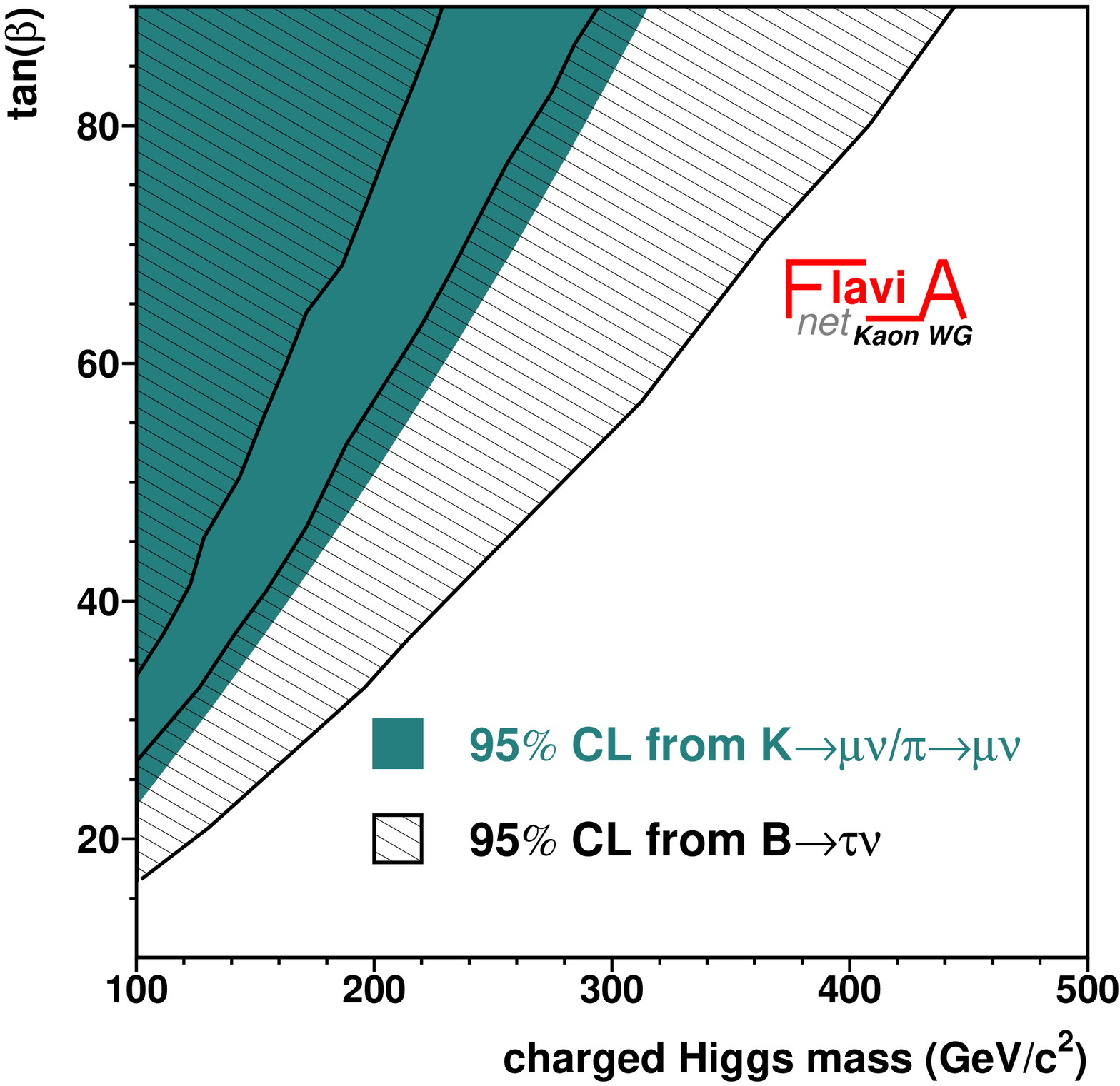} 
\caption{Present constraints in the $M_{H}$--$\tan\beta$ plane
from $\BR(\Btaun)$ and $\BR(K\to \mu\nu)$ \cite{Flavianet}.
\label{fig:Pln}}
\end{figure}

In the $B$ case only the $\tau$ modes has been observed:
${\cal B}(\Btaun)^{\rm exp} = (1.41 \pm 0.43)\times 10^{-4}$~\cite{Btaunu_exp}.
In the Kaon system the precision of
$\BR(K\to \mu\nu)$ is around $0.3\%$~\cite{KLOEkmn}.
In the limit of negligible theoretical errors, 
we should therefore expect similar bounds 
in the  $M_{H}$--$\tan\beta$ plane from 
$B$ and $K$ decays. This limit is far from being 
realistic, due to the sizable errors on $f_P$ 
(determined from Lattice QCD) and $V_{uq}$ 
(which must be determined without 
using the information on $P \to \ell \nu$ decays).
But again the present level of precision is such that 
the $B$ and $K$ decays set competitive bounds in the 
$M_{H}$--$\tan\beta$ plane (see Fig.~\ref{fig:Pln}).
Both channels have interesting possibility 
of improvement in the near future. 

\subsubsection{$B\to \ell^+\ell^-$}

The important role of $\BR(B_{s,d} \to \ell^+ \ell^-)$
in the large $\tan\beta$ regime of the MSSM  has been widely 
discussed in the literature (see 
e.g.~Ref.~\cite{Lunghi,IP,Carena:2006ai,Ellis:2007fu}
for a recent discussion).
Similarly to $P\to \ell\nu$ decays, 
the leading non-SM contribution in 
$B\to \ell^+\ell^-$ decays is generated by a
single tree-level type amplitude:
the neutral Higgs exchange 
$B\to A,H \to \ell^+\ell^-$. 
Since the effective FCNC coupling of the neutral 
Higgs bosons appears only at the quantum level, in this 
case the amplitude has a strong dependence on 
other MSSM parameters in addition to  $M_{H}$ and $\tan\beta$.
In particular, a key role is played by $\mu$ and 
the up-type trilinear soft-breaking term ($A_U$),
which control the strength of the 
non-holomorphic terms. The leading parametric 
dependence of the scalar FCNC amplitude from these 
parameters is given by 
\bea
\cA_{\rm Higgs}( B \to \ell^+\ell^-) \propto
 \frac{m_b m_\ell}{M_A^2}
 \frac{\mu A_U}{M^2_{\tilde q}} \tan^3\beta 
\times f_{\rm loop}\!\!\!\!  \no
\eea

\begin{figure}[t]
\centering
\vskip  0.3 cm
\hskip -1.5 cm
\includegraphics[scale=0.5]{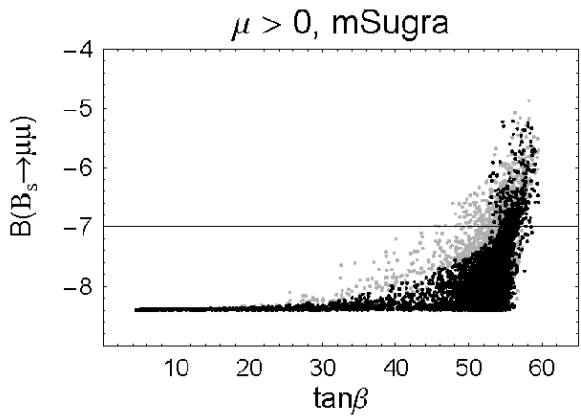} 
\vskip  0.3 cm
\caption{\label{fig:porod}  $\BR(B_s \to \mu^+\mu^-)$
as a function of $\tan\beta$ in the mSUGRA scenario~\cite{Lunghi}. }
\end{figure}

For $\tan\beta \sim 50$ and $M_A \sim 0.5$~TeV
the neutral-Higgs contribution to $\BR(B_{s,d} \to \ell^+ \ell^-)$
can easily lead to an $\cO(100)$ enhancement over the SM expectation.
This possibility is already excluded by experiments: the up\-per 
bound $\BR(B_s \to \mu^+\mu^-)< 5.8 \times 10^{-8}$~\cite{Bmm} 
is only about 15 times higher that the SM prediction
of $3.5 \times 10^{-9}$~\cite{BmmSM}.
This limit poses interesting constraints on the MSSM parameter space,
especially for light $M_H$ and large values 
of $\tan\beta$ (see e.g.~Fig.~\ref{fig:porod}). 
However, given the specific dependence on $A_U$ and $\mu$,
the present $\BR(B_s \to \mu^+\mu^-)$ bound does not exclude 
the large $\tan\beta$ effects in $(g-2)_\mu$ and 
$P\to\ell\nu$ already discussed. 
The only clear phenomenological conclusion which can be 
drawn for the present (improved) limit on $\BR(B_s \to \mu^+\mu^-)$
is the fact that the neutral-Higgs contribution 
to $\Delta M_{B_{s}}$ \cite{Buras}
is negligible.

\subsubsection{$B\to X_s \gamma$ }

The radiative decay $B\to X_s \gamma$ is one 
of the  observables most sensitive to 
non-standard contributions, not only in the large
$\tan\beta$ regime of the MSSM. 
Contrary to pure leptonic decays discussed before, 
$B\to X_s \gamma$ does not receive effective 
tree-level contributions from the Higgs sector.
The one-loop charged-Higgs amplitude, which increases the
rate compared to the SM expectation, can be partially compensated 
by the chargino-squark amplitude, giving rise to delicate cancellations.  
As a result, the extraction of bound in the $M_H$--$\tan\beta$ plane
from $\BR(B\to X_s \gamma)$ (within the MSSM) 
is a non trivial task.  

Despite the complicated interplay of various non-standard 
contributions, $B\to X_s \gamma$ is particularly interesting 
given the good theoretical control of the SM prediction
and the small experimental error. 
According to the recent NNLO analysis of Ref.~\cite{bsgth},
the SM prediction is
\bea
{\cal B}(B\to X_s \gamma)_{E_\gamma > 1.6~{\rm GeV}}^{\rm SM}
= (3.15 \pm 0.23) \times 10^{-4}  \no
\eea
to be compared with the experimental average~\cite{HFAG}:
\bea
{\cal B}(B\to X_s \gamma)_{E_\gamma > 1.6~{\rm GeV})}^{\rm exp}
 = (3.55 \pm 0.24) \times 10^{-4} \no
\eea
These results allow a small but non negligible
positive non-standard contribution to $\BR(B\to X_s \gamma)$
(as expected if the charged-Higgs amplitude would dominate over the 
chargino-squark one), which represents one of the most significant 
constraint  in the MSSM parameter space.

\begin{figure}[t]
\hskip -0.7 cm 
\includegraphics[scale=0.42]{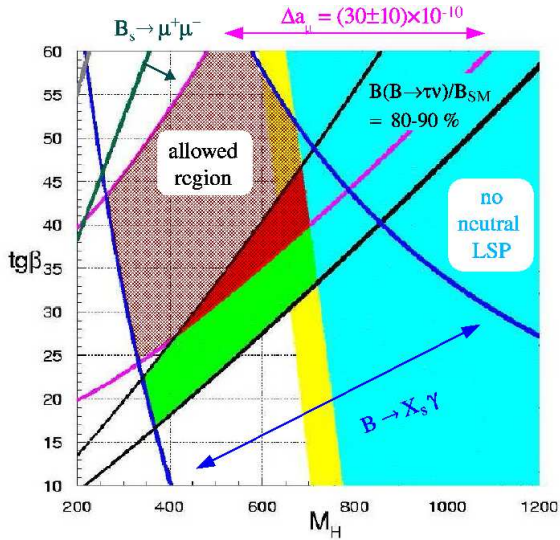} 
\caption{\label{fig:mu500}
Combined bounds from low-energy observables 
in the $\tan\beta$--$M_H$ plane assuming heavy squarks and 
dark-matter constraints in the $A$-funnel region~\protect\cite{Dark}
($M_{\tilde q} =1.5$~TeV, $A_U = -1$~TeV, $\mu=0.5$~TeV,
$M_{\tilde{\ell}}=0.4$~TeV, $1.01 < R_{ Bs\gamma} <1.24$; the 
light-blue area is excluded by the dark-matter 
conditions).
}
\end{figure}

An illustration of the typical correlations 
of the low-energy flavour constraints in the $M_H$--$\tan\beta$, 
in a generic scenario with heavy squarks and 
dark-matter condi\-tions satisfied in the 
$A$-funnel region, is shown in  Fig.~\ref{fig:mu500}.
One of the most interesting aspects of this scenario 
is the fact that a supersymmetric contribution to $a_\mu$
of $\cO(10^{-9})$ is both compatible with the present constraints 
from $\BR(B \to X_s \gamma)$ and it implies a 
suppression of  $\BR(\Btaun)$ with respect to 
its SM prediction of at least  $10\%$~\cite{Dark}.
A more precise determination of $\BR(\Btaun)$ 
is therefore a key element to test this scenario.

\subsection{Rare $K$ decays}

Among the many rare $K$ and $B$ decays, 
the $\kpn$ and $\klpn$ modes are unique since their SM branching ratios
can be computed to an exceptionally high degree of precision, not
matched by any other FCNC processes involving quarks. 
It is then not surprising that $K\to \pi \nu\bar\nu$  decays continue to 
raise a strong theoretical interest, both within 
and beyond the SM (see e.g.~Ref.~\cite{Haisch}).

\begin{figure*}[t]
  \begin{center}
   \vskip -0.4 cm
   \includegraphics[width=0.85\linewidth]{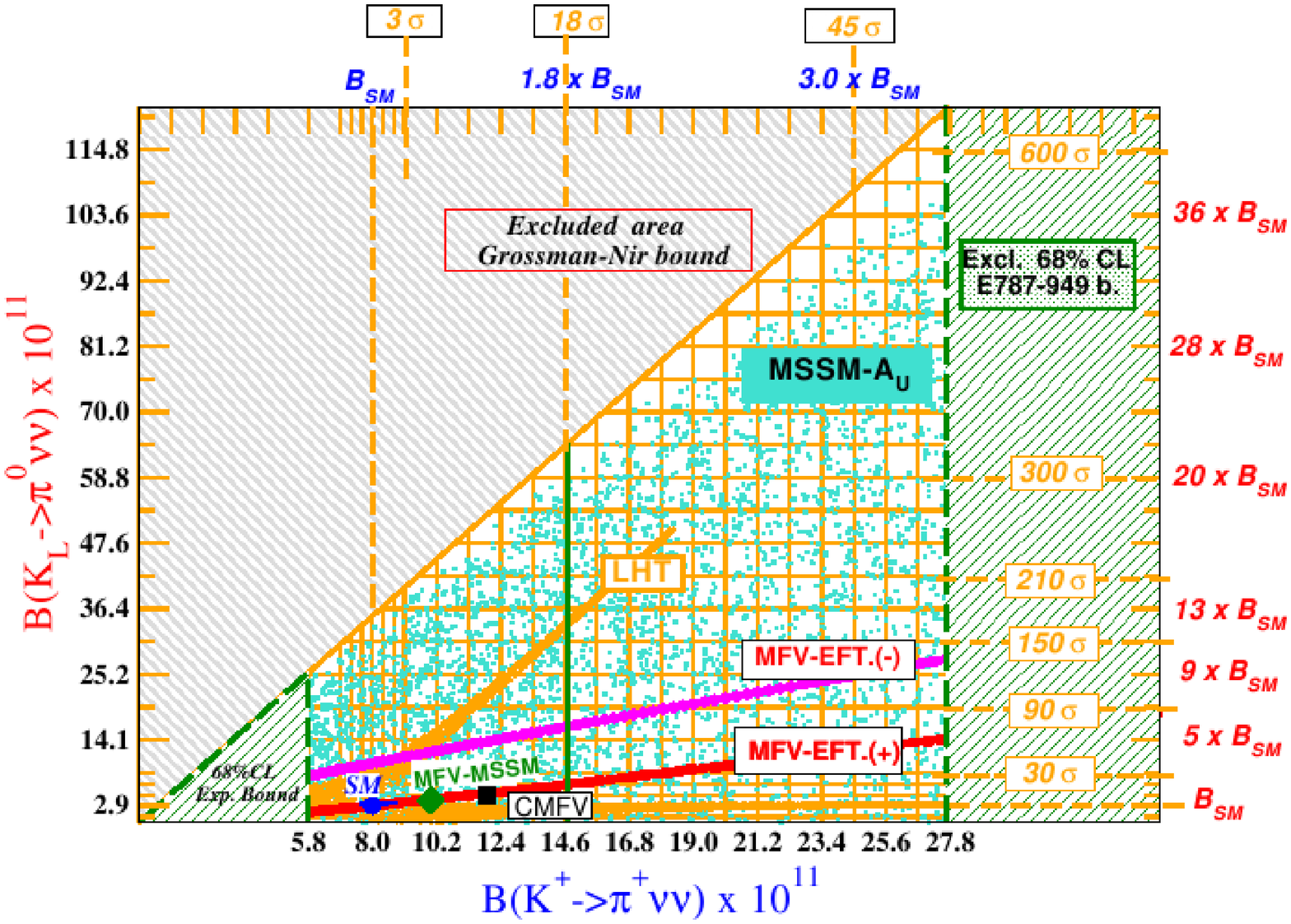}
   \vskip -0.8 cm
  \end{center}
  \caption{Predictions of different NP models for $\BR(\kpn)$
and $\BR(\kpn)$  [courtesy of F.~Mescia]. The 95\%~C.L.~exluded areas of 
$\BR(\kpn)$ refer to the result of the BNL-E787/949 
experiment~\cite{E949}. \label{fig:Mesciaplot}}
\end{figure*}

Because of the strong suppression of 
the $s \to d$ short-distance amplitude in the SM [$V_{td}V_{ts}^* =\cO(10^{-4})$],
rare $K$ decays are the most sensitive probes of possible deviations from the  
strict MFV ansatz. 
Several recent NP analyses confirm the high discovery potential 
of these channels (see Fig.~\ref{fig:Mesciaplot} and Ref.~~\cite{Haisch}).
The latter has also improved thanks three significant improvements 
on the SM predictions of $K\to \pi \nu\bar\nu$ rates: i) the NNLO calculation 
of the dimension-six charm-quark contribution to $\kpn$~\cite{NNLL}; 
ii) the first complete analysis of dimension-eight and long-distance 
(up-quark) contributions relevant to $\kpn$~\cite{IMS}; 
iii) a new comprehensive analysis of matrix-elements and isospin-breaking effects,
relevant to both channels~\cite{MS}.
Thanks to these recent works, the irreducible theoretical 
uncertainties on both branching ratios are at the few \% level.

It is worth stressing that if a deviation from the SM 
is seen in one of the two $K\to\pi \nu\bar \nu$ channels, a key 
independent information about the nature of NP can be obtained 
also from the two $K_L \to \pi^0 \ell^+\ell^-$ ($\ell=e,\mu$) modes. 
The latter are not as clean as the neutrino modes, but are still dominated by 
SD dynamics and very sensitive to 
NP~\cite{Buchalla:2003sj,Isidori:2004rb,Prades,Mescia:2006jd}.

\subsection{Lepton Flavour Violation and LF non-universality}
\label{sect:LFV}

LFV couplings naturally appear in the MSSM once we extend 
it to accommodate the non-vanishing neutrino masses 
and mixing angles by means of a supersymmetric seesaw mechanism~\cite{fbam}.
In particular, the renormalization-group-induced LFV entries 
appearing in the left-handed slepton mass matrices 
have the following form~\cite{fbam}:
$\delta_{LL}^{ij} = c_\nu (Y^\dagger_\nu Y_\nu)_{ij}$,
where  $Y_\nu$ are the neutrino Yukawa couplings
and $c_\nu$ is a numerical coefficient of  $\cO(0.1$--$1)$.
The information from neutrino 
masses is not sufficient to determine in a model-independent 
way all the seesaw parameters relevant to LFV rates and,
in particular, the neutrino Yukawa couplings. 
To reduce the number of free parameters specific SUSY-GUT 
models and/or flavour symmetries need to be employed.
Two main roads are often considered in the literature:
the case where the charged-lepton LFV couplings are linked 
to the CKM matrix (the quark mixing matrix) and the case where 
they are connected to the PMNS matrix (the neutrino mixing 
matrix)~\cite{Arganda:2005ji}.

Once non-vanishing LFV entries in the slepton mass matrices 
are generated, LFV rare decays are naturally induced by
one-loop diagrams with the exchange of gauginos and sleptons.
For large values of $\tan\beta$ the radiative decays
$\ell_{i}\rightarrow\ell_{j}\gamma$, mediated by 
dipole operators, are linearly enhanced, in close analogy to 
the  $\tan\beta$-enhancement of $\Delta a_\mu=(g_\mu-g^{\rm SM}_\mu)/2$.
A strong link between 
these two observable naturally emerges~\cite{hisano}.
We can indeed write
\bea
&& \frac{\BR(\ell_i\rightarrow \ell_j\gamma)}
{\BR(\ell_i\rightarrow \ell_j\nu_{\ell_i}\bar{\nu_{\ell_j}})} 
= \frac{48\pi^{3}\alpha}{G_{F}^{2}}
\left[\frac{\Delta a_{\mu}}{m_{\mu}^{2}}\right]^{2} \times \no \\ 
&& \qquad \times \left[\frac{f_{2c}\left( M^{2}_{2}/M^{2}_{\tilde \ell}, \mu^2/M^{2}_{\tilde \ell} \right)}{
g_{2c}\left( M^{2}_{2}/M^{2}_{\tilde \ell}, \mu^2/M^{2}_{\tilde \ell} \right)}
\right]^2
\,\left| \delta_{LL}^{ij} \right|^2 \qquad
\label{eq:ratio_LFV}
\eea
where $f_{2c}$ and $g_{2c}$ are $\cO(1)$ loop functions. 
In the limit of a degenerate SUSY spectrum, this implies 
\bea
&& \BR(\ell_i\rightarrow \ell_j \gamma) ~\approx~ 
\left[\frac{\Delta a_{\mu}}{ 20 \times 10^{-10}}\right]^{2} \times \no \\
&& \qquad \times
 \left\{
\ba{ll}
1 \times 10^{-4} \, \left| \delta_{LL}^{12} \right|^2 \qquad  & [\mu\to e]  \\
2 \times 10^{-5} \, \left| \delta_{LL}^{23} \right|^2         & [\tau\to \mu]
\ea \qquad 
\right.
\label{eq:corr}
\eea

\begin{figure}[t]
\centering
\includegraphics[scale=0.42]{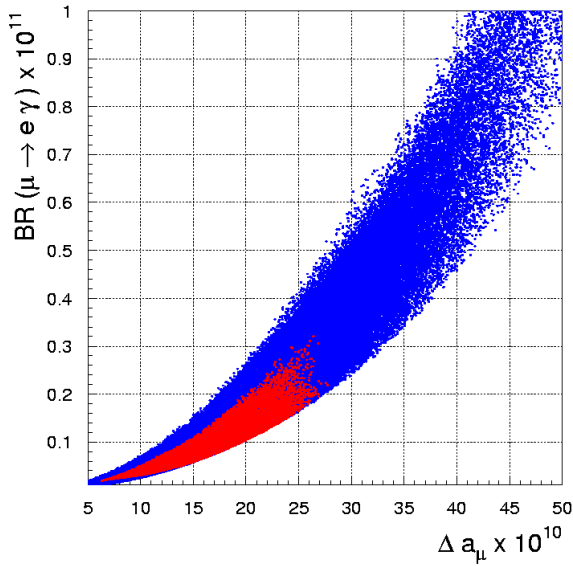} 
\caption{\label{fig2}
 $\cB(\mu\rightarrow e\gamma)$  
vs.~$\Delta a_\mu=(g_\mu-g^{\rm SM}_\mu)/2$ in the MSSM
assuming $|\delta_{LL}^{12}|=10^{-4}$~\protect\cite{IP}. }
\end{figure}

The strong correlation between $\Delta a_{\mu}$ and the 
rate of the LFV transitions $\ell_i \rightarrow \ell_j\gamma$
holds well beyond the simplified assumptions used to 
derive these equations (see Fig.~\ref{fig2}).
The normalization $|\delta_{LL}^{12}|=10^{-4}$ used in 
Fig.~\ref{fig2} for  $\cB(\mu\rightarrow e\gamma)$
corresponds to the MFV hypothesis in the 
lepton sector with $M_\nu \gsim 10^{12}$ GeV~\cite{MLFV}. 
As can be seen, for such natural choice of $\delta_{LL}$ 
the $\mu\rightarrow e\gamma$ branching ratio is in the $10^{-12}$ 
range, i.e.~well within the reach of the MEG experiment~\cite{MEG}.

Ratios of similar LFV decay rates, 
such as $\mathcal{B}(\tau\to \mu\gamma)/\mathcal{B}(\mu\to e\gamma)$, 
are much more easy to be predicted, being 
free from the overall normalization uncertainty.
These predictions depend essentially only on the flavour 
structure of the LFV couplings. The search for 
$\mathcal{B}(\tau\to \mu\gamma)$ is thus a key element 
in trying to determine the structure of flavour symmetry 
breaking in the lepton sector. In particular,
$\mathcal{B}(\tau\to \mu\gamma)/\mathcal{B}(\mu\to e\gamma)$
ranges from to $10^2$ in the case of a PMNS hierarchy,
to $10^4$ in the case of a CKM-type hierarchy.
In the latter case $\mathcal{B}(\tau\to \mu\gamma)$ 
can exceed $10^{-9}$ and be within the reach of a super-$B$ factory. 
The enhancement of
$\mathcal{B}(\tau\to \mu\gamma)$ can be even larger  
in non-supersymmetric frameworks, such as 
the one recently discussed in Ref.~\cite{PQ}.

An independent and potentially large class of LFV 
contributions to rare decays in the 
large $\tan\beta$ regime of the MSSM 
comes from Higgs-mediated amplitudes. 
Similarly to the quark sector, 
non-holomorphic couplings can induce an effective FCNC 
Higgs coupling also in the 
lepton sector~\cite{Babu:2002et}.
Gauge- and Higgs-me\-dia\-ted amplitudes 
leads to very different 
correlations among 
LFV processes~\cite{Arganda:2005ji,Paradisi:2005tk,Paradisi:2006jp}
and their combined study can reveal 
the underlying mechanism of LFV.

Finally, as recently pointed out in Ref.~\cite{Masiero:2005wr},
Higgs-mediated LFV effects at large $\tan\beta$ can also induce visible 
deviations of lepton-flavour universality 
in char\-ged-current processes. 
If the slepton sector contains 
sizable (non-minimal) sources of LFV, we could hope to observe 
deviations from the SM predictions in the 
$\BR(P\to \ell \nu)/\BR(P \to \ell^\prime \nu)$
ratios. The deviations can be $\cO(1\%)$ 
in $\BR(K\to e \nu)/\BR(K \to \mu \nu)$ \cite{Masiero:2005wr}, 
and  can reach $\cO(1)$ and $\cO(10^3)$ in  
$\BR(B\to \mu \nu)/\BR(B \to \tau \nu)$ and 
$\BR(B\to  e \nu)/\BR(B \to \tau \nu)$, respectively~\cite{IP}.

\subsection{Other observables}

The observables mentioned so far are only a subset of those which 
is still worth to improve, or to search for, in the LHC era. They 
have been selected mainly because they are interesting also in MFV 
scenarios, or in the most pessimistic case for flavour physics. 
Going beyond MFV --as stressed above, this possibility is 
certainly open-- the list of potentially interesting measurements 
is much longer. Extensive studies can be found in the recent 
reports~\cite{Buchalla:2008jp,Raidal:2008jk,Bona:2007qt}.
Mentioning only a few examples, a key observable to falsify 
MFV is the time-dependent CP asymmetry in $B_s \to \psi \phi$
(or the measurement of the CPV phase of $B_s$--$\bar B_s$ 
mixing), one of the golden channels of the LHCb experiment. 
It is also very important, both within and beyond MFV,
trying to improve the measurements of CKM elements 
from tree-level dominated processes (namely $|V_{us}|$ from 
$K_{\ell 3}$, $|V_{ub}|$ and $|V_{cb}|$ from semileptonic $B$ 
decays, and $\gamma$ from CP asymmetries in penguin-free modes).
These measurements are not directly sensitive to NP, 
but are the key ingredient to improve the SM predictions 
in processes which are sensitive to NP.  

On general grounds, a key issue in planning 
future experiments in the quark sector 
is the control of theoretical uncertainties, or the control
over long-distance dynamics. From this point of view, leptonic and semileptonic 
$B$ and $K$ decays (both charged and neutral currents, either fully 
inclusive or with at most one stable meson in the final state),
are the potentially most interesting channels. 
In the $B$ case we should also add CP asymmetries in penguin-free 
modes and some radiative decays (most notably $B\to X_s \gamma$). 
On the other hand, some of the observables 
which have received a lot of attention in the recent past,
such as time-dependent CP asymmetries in penguin dominated 
modes, are less interesting: the present level of accuracy 
is not far from the level of irreducible theoretical 
uncertainties, and no sizable deviation from the SM has 
been identified yet. 

Last but not least, we comment about $D$--$\bar D$ mixing,
whose evidence reported by $B$-factory experiments 
is one of the highlights of this conference~\cite{DD1}. 
Such evidence is a very useful information 
about the interplay of weak and strong interactions.
However, the impact about physics 
beyond the SM is rather limited. The observed values of 
the $D$--$\bar D$ mixing parameters are in the ball park 
of the SM expectations. This allows us to exclude 
models which predict a too large $\Delta m_D$ (see e.g.~\cite{DD2}), 
but the bounds are not very precise given the 
sizable long-distance contributions to this quantity.
In a future perspective, only the measurement 
of CP violation in $D$--$\bar D$ mixing could 
provide a significant new information about physics 
beyond the SM.

\section{Conclusions} 
 
The absence of significant deviations from the SM 
in quark flavour physics is a key information about 
new physics. Only models with highly non 
generic flavour structures can both stabilise the 
electroweak sector and be compatible with flavour 
observables. In such models we expect new particles 
within the LHC reach. 
This does not mean that the complete
structure of the new theory can easily be determined at the LHC.
Virtually in all cases interesting aspects of the models
can be determined only from future high-precision studies 
in the flavour sector. 

As briefly illustrated in this talk, 
the set of low-energy observables to be measured 
with higher precision, and the rare transitions to be searched for 
is limited (if we are interested only on physics beyond the SM). 
However, there is not a single experimental set up where 
we can study all of them.
The set of interesting observables includes 
$\mu$, $K$, $\tau$, and $B$ decays. Ideally, their 
precision study would require several 
different types of facilities 
and experiments.

\section*{Acknowledgments} 
It is a pleasure to thank the organizers of Lepton Photon 2007 
for the invitation to present this talk.
This work has been supported in part by the EU Contract No.
MRTN-CT-2006-035482, ``FLAVIAnet''.

\end{document}